\documentclass{jpp}
\usepackage{amsmath,amssymb,bm}
\usepackage{graphicx,color}
\pdfoutput=1

\usepackage[utf8]{inputenc}
\usepackage[T1]{fontenc}

\shorttitle{Whistler precursor}
\shortauthor{G. Granit and M. Gedalin}

\title{Whistler precursor and intrinsic variability of quasi-perpendicular shocks}

\author{Gilad Granit\aff{1}
 \and Michael Gedalin\aff{1} \corresp{\email{gedalin@bgu.ac.il}},}

\affiliation{\aff{1}Department of Physics, Ben-Gurion University of the Negev, Beer-Sheva, Israel}

\DeclareMathOperator\real{Re}
\begin{document}

\maketitle

\begin{abstract}
The structure of whistler precursor in a quasi-perpendicular shock is studied within two-fluid approach in one-dimensional case. The complete set of equations is reduced to the KdV equation, if no dissipation is included. With a phenomenological resistive dissipation the structure is described with the KdV-Burgers equation. The shock profile is intrinsically time dependent. 
For sufficiently strong dissipation, temporal evolution of a steepening profile results in generation of  a stationary decaying whistler ahead of the shock front. With the decrease of the dissipation parameter  whistler wavetrains begin to detach and propagate toward upstream and the ramp is weakly time dependent.   
 In the weakly dissipative regime the shock front experiences reformation. 
\end{abstract}

\section{Introduction}
Collisionless shocks, even at low-Mach numbers, are complicated structures. The majority of the shocks, encountered in the heliosphere, are fast magnetized shocks, where the shock transition is mainly determined   
by the rapid increase of the magnetic field \citep[see, e.g.][ for a review]{Kennel1985A-quarter-centu}. While the magnetic jump in shocks must be accompanied by the drop of the fluid speed from super-magnetosonic to sub-magnetosonic, by the jump in the ion and electron density, and by the substantial increase of temperatures of both species, it is the magnetic field which is a) measured with the highest precision, and b) serves as the main signature of the shock. Knowledge of the structure of the magnetic profile is of utmost importance, since the macroscopic magnetic field is the most important factor affecting the particle motion throughout the shock. The most important parameters defining the shock behavior are thought to be the Mach number $M$ (defined below), the angle $\theta$ between the shock normal and the upstream magnetic field, and the  ratio of the upstream kinetic-to-magnetic pressure $\beta=8\pi p_u/B_u^2$. Here $p_u=n_uT_u$ is the upstream plasma pressure and $B_u$ is the upstream magnetic field. Hereafter the subscript $u$ refers to the upstream values and $d$ refers to the downstream parameters. Low-Mach number low-$\beta$ shocks are believed to be laminar, that is, to possess an almost monotonic increase of the magnetic field with nearly no magnetic oscillations \citep{Greenstadt1975Structure-of-th,Mellott1984The-structure-o,Farris1993Magnetic-struct}. Recently, it was shown that even in very low-Mach number shocks downstream magnetic oscillations are caused by ion gyration \citep{Balikhin2008Venus-Express-o,Russell2009STEREO-observat,Ofman2009Collisionless-r,Kajdic2012Waves-upstream-,Ofman2013Two-dimensional,Gedalin2015Collisionless-r}. Low-Mach number shocks also often exhibit magnetic oscillations upstream of the ramp \citep{Balikhin2003Identification-,Blanco-Cano2006Macrostructure-,Wilson2007Waves-in-Interp,Wilson2009Low-frequency-w,Ramirez-Velez2012Whistler-waves-,Blanco-Cano2013STEREO-interpla,Blanco-Cano2016Interplanetary-,Wilson-III2017Revisiting-the-}. The latter are typically identified as low-frequency electromagnetic whistlers \citep{Ramirez-Velez2012Whistler-waves-}. Whistler generation is quite typical for supercritical shocks \citep{Walker1999Ramp-nonstation,Scholer2007Whistler-waves-,Wilson2009Low-frequency-w,Hull2012Multiscale-whis}. Often these whistlers propagate at an angle to the shock normal and possibly precede the formation of ripples in supercritical shocks \citep{Lowe2003The-properties-,Burgess2007Shock-front-ins,Lobzin2008On-nonstationar,Johlander2016Rippled-Quasipe}. It was also suggested that at higher Mach numbers whistler generation evolves into strong non-stationarity of the shock front, known as reformation \citep{Walker1999Ramp-nonstation,Krasnoselskikh2002Nonstationarity,Scholer2007Whistler-waves-,Lobzin2008On-nonstationar,Lembege2009Nonstationarity,Krasnoselskikh2013The-Dynamic-Qua,Sulaiman2015Quasiperpendicu}. It is general understanding that stationarity of even low-Mach number shocks requires presence of some dissipation, presumably in the form of resistivity. The de facto standard semi-qualitative description of the shock front suggests transformation of a soliton solution into a shock-like profile when dispersion is accompanied with weak dissipation \citep{Sagdeev1966Cooperative-Phe}. It was shown that in the presence of weak resistive dissipation shock-like solutions do exist in one-dimensional two-fluid quasineutral approximation \citep{Gedalin1998Low-frequency-n}, while without dissipation there are no solutions with different asymptotic values of the magnetic field. Appearance of the downstream magnetic oscillations is a kinetic effect \citep{Ofman2009Collisionless-r,Gedalin2015Collisionless-r} and cannot be properly described within the widely accepted single polytropic two-fluid theory. Yet, upstream of the ramp the two-fluid approximation is still valid. It was shown that even if restricting for the upstream region only, stationary solutions with decaying oscillations exist only if dissipation is invoked \citep{Gedalin2015Profile-of-a-lo}. This brings us to the conclusion that even low-Mach shocks may be intrinsically nonstationary and  generate upstream propagating whistler wavetrains. Weakly nonlinear weakly dispersive waves are typically described by the Korteveg-de-Vries equation (KdV) \citep{Karpman1975Non-linear-wave,Ablowitz2011Nonlinear-Dispe}. Asymptotic behavior of the initial discontinuity in KdV has been studied using the Witham method \citep{Gurevich1974Nonstationary-s} and by direct numerical integration \citep{Ablowitz2009Soliton-generat,Ablowitz2013Dispersive-shoc}. However, in the physically meaningful situation an initially smooth profile steepens and wavetrain generation is expected to prevent discontinuity formation. The physically interesting stage, to be compared with observations, is not the asymptotic state but the transient one, taking also into account the variability of the ambient conditions. In general, one-dimensional time-dependent nonlinear  waves in two-fluid theory are described by a set of equations which are more complicated than KdV \citep{Gedalin1993Nonlinear-waves,Gedalin1994Nonlinear-waves}. In the present paper we derive the time-dependent equations for one-dimensional waves in two-fluid theory with resistive dissipation included. We reduce these equations to a single KdV-Burgers equation and analyze it numerically. We show that in the case of sufficiently weak dissipation an initially smooth profile with different asymptotic steepens to an intrinsically non-stationary shock generating upstream wavetrains.   

\section{Basic equations}

The analysis will be done within two-fluid plasma theory with massless electrons, since the typical spatial scales are much larger than the electron inertial length and the typical temporal scales are much larger than the electron gyroperiod or inverse electron plasma frequency. For the same reason  quasineutrality, $n_e=n_i=n$, is maintained with high precision. The speeds are well below the speed-of-light, so that the displacement current is negligible. 
We also assume that the system is one-dimensional, that is, all plasma variables  and fields depend only on the coordinate $x$ along the shock normal and on time $t$. With these approximations the two-fluid equations
take the form
\begin{align}
&\partial_t n+\partial_x(nv_x)=0\label{eq:density}\\
&nm_i\left(\partial_t v_x+v_x\partial_x v_x\right)=ne
\left(E_x+\bm{\hat{x}}\cdot (\bm{v}_i\times \bm{B})/c\right) - \partial_xp_i\label{eq:ivx}\\
&0=-en
\left(E_x+\bm{\hat{x}}\cdot (\bm{v}_e\times \bm{B})/c\right) - \partial_xp_e\label{eq:evx}\\
&nm_i\left(\partial_t \bm{v}_{i\perp}+v_x\partial_x \bm{v}_{i\perp}\right)=ne
\left(\bm{E}_\perp+v_x\bm{\hat{x}}\times \bm{B}_\perp/c+B_x \bm{v}_{i\perp}\times \bm{\hat{x}}/c\right)-n\nu(\bm{v}_{i\perp}-\bm{v}_{e\perp})\label{eq:ivperp}\\
&0=-en
\left(\bm{E}_\perp+v_x\bm{\hat{x}}\times \bm{B}_\perp/c+B_x \bm{v}_{e\perp}\times \bm{\hat{x}}/c\right)-n\nu(\bm{v}_{e\perp}-\bm{v}_{i\perp})\label{eq:evperp}\\
&c\partial_x(\bm{\hat{x}}\times\bm{B}_\perp)=4\pi ne(\bm{v}_{i\perp}-\bm{v}_{e\perp})\label{eq:ampere}\\
&\partial_t\bm{B}_\perp=-c\partial_x(\bm{\hat{x}}\times \bm{E}_\perp)\label{eq:induction}\\
&B_x=\text{const}, \quad v_{ex}=v_{ix}=v_x\label{eq:bxvx}
\end{align}
Here $n$, $v_x$, and $\bm{v}_\perp$ are the ion density and velocity, $B_x$ and $\bm{B}_\perp$ are the magnetic field components, while $E_x$ and $\bm{E}_\perp$ are the electric field components. The subscript $\perp$ denotes components perpendicular to the shock normal: $\bm{v}_\perp=(v_y,v_z)$, $\bm{B}_\perp=(B_y,B_z)$, $\bm{E}_\perp=(E_y,E_z)$.
Here $\perp$ refers to the direction $\bm{\hat{x}}$. We shall assume that $p_e$ and $p_i$ are functions of the density only. The last terms in \eqref{eq:ivperp} and \eqref{eq:evperp} describe the friction (momentum exchange) between the two species which conserves the total momentum. Indeed, the following conservation laws can be easily derived:  
\begin{align}
&\partial_t n +\partial_x (nv_x)=0\label{eq:numcon}\\
&\partial_t(nm_iv_x) + \partial_x\left(nm_iv_x^2+p+\frac{B_\perp^2}{8\pi}\right)=0\label{eq:pxcon}\\
&\partial_t(nm_i\bm{v}_{i\perp})+\partial_x\left(nm_iv_x\bm{v}_{i\perp}-\frac{B_x\bm{B}_\perp}{4\pi}\right)=0\label{eq:pperpcon}
\end{align} 
which are the particle number  and the momentum conservation and $p=p_i+p_e$.
Eqs. \eqref{eq:evperp} and \eqref{eq:ampere} give
\begin{align}
&c(\bm{\hat{x}}\times\bm{E}_\perp)=v_x\bm{B}_\perp -B_x\bm{v}_{i\perp} +\frac{cB_x}{4\pi ne} \partial_x(\bm{\hat{x}}\times\bm{B}_\perp) -\frac{c^2\nu}{4\pi ne^2} \partial_x\bm{B}_\perp\label{eq:xcrosse}
\end{align}
so that \eqref{eq:induction} takes the form
\begin{align}
&\partial_t\bm{B}_\perp+\partial_x(v_x\bm{B}_\perp -B_x\bm{v}_{i\perp} ) +\partial_x\left(\frac{cB_x}{4\pi ne} \partial_x(\bm{\hat{x}}\times\bm{B}_\perp) -\frac{c^2\nu}{4\pi ne^2} \partial_x\bm{B}_\perp\right)=0\label{eq:induction1}
\end{align}
Let us choose some point as a  reference point. It is always possible to switch to a reference frame such that in the reference point $\bm{v}_{i\perp}=0$. It is always possible to rotate to a coordinate system around axis $x$ in such  a way so that in the reference point $B_y=0$. We shall denote the other variables in the reference point as follows: $n=n_0$, $B_x=B_0\cos\theta$,   $B_z=B_0\sin\theta$, and $p=p_0$.

We further define 
\begin{align}
&v_A^2=\frac{B_0^2}{4\pi n_0m_i}, \quad
\omega_{pi}^2=\frac{4\pi n_0e^2}{m_i}, \quad \Omega=\frac{eB_0}{m_ic}, \quad\beta=\frac{8\pi p_0}{B_0^2}, \quad M=\frac{v_0}{v_A}
\end{align}
It is convenient to introduce the normalized  variables as follows:
\begin{align}
& T=\Omega t, \quad X=\frac{\omega_{pi} x}{c},\quad N=\frac{n}{n_0}, \quad v=\frac{v_x}{v_A}, \\
& \bm{u}=\frac{\bm{v}_{i\perp}}{v_A}, \quad \bm{b}=\frac{\bm{B}_\perp}{B_0}
\end{align}
We shall also assume the polytropic state equation $p/n^\gamma=\text{const}$.
In the normalized form the equations are written as follows:
\begin{align}
&\partial_T N+\partial_X(Nv)=0\label{eq:normdens2}\\
&\partial_T(Nv)+\partial_X\left(Nv^2+\frac{\beta N^\gamma+b^2}{2}\right)=0\label{eq:normvx2}\\
&\partial_T(N\bm{u})+\partial_X\left(Nv\bm{u}-\cos\theta\bm{b}\right)=0\label{eq:normu2}\\
&\partial_T\bm{b}+\partial_X(v\bm{b}-\cos\theta\bm{u}) \notag\\
&+\cos\theta \partial_X(N^{-1}\partial_X(\bm{\hat{x}}\times\bm{b}))-\mu\partial_X(N^{-1}\partial_X\bm{b})=0\label{eq:normb2}
\end{align}
It is  useful to write down the above equations using the variables $J=Nv$, and $\bm{W}=n\bm{u}$:
\begin{align}
&\partial_T N+\partial_X J=0\label{eq:normdens3}\\
&\partial_T J+\partial_X\left(\frac{J^2}{N}+\frac{\beta N^\gamma+b^2}{2}\right)=0\label{eq:normvx3}\\
&\partial_T\bm{W}+\partial_X\left(\frac{\bm{W}}{N}-\cos\theta\bm{b}\right)=0\label{eq:normu3}
\end{align}

\section{Stationary states}

If there is no time dependence we arrive at the following conservation laws
\begin{align}
&J=M\label{eq:normdens5}\\
&\frac{M^2}{N}+\frac{\beta N^\gamma +b_y^2+b_z^2}{2}
=M^2+\frac{\beta+\sin^2\theta}{2}\label{eq:normvx5}\\
&\frac{W_y}{N}-\cos\theta b_y=0\label{eq:normu5y}\\
&\frac{W_z}{N}-\cos\theta b_z=-\cos\theta\sin\theta \label{eq:normu5z}
\end{align}
while the remaining equations for $b_y$ and $b_z$ take the form
\begin{align}
&Mb_y-\cos\theta W_y-\cos\theta \partial_Xb_z-\mu\partial_Xb_y=C_y\label{eq:normb5y}\\
&Mb_z-\cos\theta W_z+\cos\theta \partial_Xb_y-\mu\partial_Xb_z=C_z\label{eq:normb5z}
\end{align}
where the constants $C_y$ and $C_z$ should be determined by additional conditions in the reference point. Since we are particularly interested in  solutions which asymptotically converge to an  uniform state, we take the reference point at $x\rightarrow \pm\infty$ and assume that all derivatives vanish there, that is, $C_y=0$ and $C_z=\sin\theta$. Eventually, one has
\begin{align}
&\frac{1}{N}+\frac{\beta N^\gamma}{2M^2}+\frac{b^2}{2M^2}=1+\frac{\beta+\sin^2\theta}{2M^2}\label{eq:P2}\\
&b_y(M^2-N\cos^2\theta)-\cos\theta \partial_Xb_z-\mu\partial_Xb_y=0\label{eq:by2}\\
&b_z(M^2-N\cos^2\theta)+\cos\theta\partial_Xb_y-\mu\partial_Xb_z=N\sin\theta(M^2-\cos^2\theta)\label{eq:bz2}
\end{align}
We start with the analysis of the stationary point by linearizing the equations for   $N=1+\delta N$, $b_z=\sin\theta+\delta b_z$, $b_y=\delta b_y$ and searching for solutions $\propto \exp(kx)$~\citep[see, e.g.][]{Gedalin1998Low-frequency-n}.  One has
\begin{align}
&(M^2-\cos^2\theta-k\mu)[(M^2-\cos^2\theta-k\mu)(M^2-\beta\gamma/2)\notag\\
&-M^2\sin^2\theta(M^2-\cos^2\theta)]+k^2\cos^2\theta(M^2-\beta\gamma)=0\label{eq:disp}
\end{align}
It is easy to recognize the MHD speeds (normalized on the Alfven speed):
\begin{align}
&v_I^2=\cos^2\theta, \quad  v_s^2=\beta\gamma/2, \quad
v_{F,SL}^2=\frac{1}{2}[(1+v_s^2)\pm\sqrt{(1+v_s^2)^2-4v_s^2\cos^2\theta}]\\
& v_F^2>v_I^2 >v_{SL}^2, \qquad  v_F^2> v_s^2 >v_{SL}^2
\end{align}
then \eqref{eq:disp} takes the form
\begin{align}
&(\cos^2\theta+\mu^2)k^2-k\mu L_1 +L_2=0\\
&L_1=\frac{(M^2-v_I^2)(M^2-v_s^2)+(M^2-v_F^2)(M^2-v_{SL}^2)}{M^2-v_s^2}\\
&L_2= \frac{(M^2-v_I^2)(M^2-v_{SL}^2)(M^2-v_F^2)}{M^2-v_s^2}\\
&k_{1,2}=\frac{\mu L_1\pm\sqrt{\mu^2L_1^2-4(\cos^2\theta +\mu^2)L_2}}{2(\cos^2\theta+\mu^2)}
\end{align}

A shock profile can exist if  the upstream point has an  unstable solution and the downstream point has a stable solution, that is, if there exist $\real k>0$ for $x\rightarrow -\infty$ (upstream) and $\real k<0$ for $x\rightarrow \infty$ (downstream). For a fast shock, in the upstream point one has $M^2>v_F^2$ and, therefore, $L_1>0$, $L_2>0$.  As a result, $\real k_{1,2}>0$, which means that the upstream point is unstable. If $\mu^2L_1^2-4(\cos^2\theta +\mu^2)L_2>0$ this point is an unstable node, otherwise it is an unstable focus (spiral point). In the downstream asymptotic point of a fast shock one has $v_F^2>M^2>v_I^2$. If $M^2<v_s^2$ then again $L_1>0$ and $L_2>0$, so that $\real k_{1,2}>0$, which means that no solution can converge to the downstream point for $M^2<v_s^2$. If $M^2>v_s^2$, then $L_2<0$ and $\real k_1>0$, $\real k_2<0$. The downstream point becomes a saddle in this case, having exactly one direction along which the solution is converging to the asymptotic state. 

In the MHD approximation $k^2$ is neglected and one has
$k\mu=L_2/L_1$, which gives an unstable upstream and stable downstream for $M^2>v_s^2$, corresponding to the classical first sub- to super-critical  transition~\citep[see, e.g.][]{Kennel1985A-quarter-centu}. From the above analysis it is clear that the MHD description loses the whistler mode and forces linear polarization throughout.

The stationary point analysis does not, however, provide all the information about the topology of the solutions, since \eqref{eq:P2} limits the magnetic field magnitude from above:
\begin{align}
&b^2_{max}=2M^2(1-N_c^{-1})+\beta(1-N_c^\gamma)+\sin^2\theta\\
&N_c=\left(\frac{2M^2}{\beta\gamma}\right)^{1/(\gamma+1)}
\end{align}

\section{Weakly nonlinear weakly time-dependent solutions}
Let us now extend our analysis onto the weakly time-dependent regime. 
In the  weakly nonlinear case the time-derivative term, the second- and third-order spatial derivative terms, and the nonlinearity due to $N$ are all small in  \eqref{eq:normu2}-\eqref{eq:normb2}. Thus, it is sufficient to express $J$, and $N$, and $\bm{W}$  from the stationary equations:
\begin{align}
& J=M=\text{const}\label{eq:normdens4}\\
&\frac{M^2}{N}+\frac{\beta N^\gamma+b^2}{2}=\text{const}\label{eq:normvx4}\\
&W_y=\cos\theta Nb_y\\
&W_z=\cos\theta N(b_z-\sin\theta)\\
&\partial_Tb_y+\partial_X \left(\frac{M^2b_y}{N}-\cos^2\theta b_y\right) -\cos\theta \partial^2_X b_z-\mu\partial^2_X b_y=0\label{eq:normby4}\\
&\partial_Tb_z+\partial_X \left(\frac{M^2b_z}{N}-\cos^2\theta b_z\right) +\cos\theta \partial^2_X b_y-\mu\partial^2_X b_z=0\label{eq:normbz4}
\end{align}
Defining $V=1/N$ and expanding further one gets 
\begin{align}
&\partial_Tb_z+\partial_X \left(VMb_z-\cos^2\theta b_z\right) +\cos\theta \partial^2_X b_y-\mu\partial^2_X b_z=0\\
&b_y=\left(\frac{\cos\theta}{VM^2-\cos^2\theta}\right)\partial_X b_z\\
&b_y\approx\left(\frac{\cos\theta}{M-\cos^2\theta}\right)\partial_X b_z\\
&\partial_Tb_z+\partial_X \left(VM^2b_z-\cos^2\theta b_z\right) + 
\left(\frac{\cos^2\theta}{M-\cos^2\theta}\right)\partial^3_X b_z-\mu\partial^2_X b_z=0\\
&(M^2 -v_s^2)(V-1)=-\sin\theta(b_z-\sin\theta)
\end{align}
After some straightforward algebra, eventually we arrive at the following equation for 
 $\xi=b_z/\sin\theta-1$:
%
\begin{align}
&\partial_T\xi+P\partial_X \xi - Q\partial_X\xi^2+ R\partial^3_X\xi - \mu\partial^2_X\xi=0\label{eq:KdVB1}\\
&P=\frac{(M^2-v_F^2)(M^2-v_{SL}^2)}{M^2-v_s^2} \\
&Q=\frac{\sin^2\theta}{M^2-v_s^2}\\
&R=\frac{\cos^2\theta}{M^2-\cos^2\theta}
\end{align}
In the stationary case one has
\begin{align}
&P\xi - Q\xi^2+ R\partial^2_X\xi - S\partial_X\xi=0
\end{align}
Let $g=\tanh kX$, so that $\partial_Xg=k(1-g^2)$.
Ignoring dispersion $R$ one gets
\begin{align}
&\partial_X\xi=\left(\frac{Q}{S}\right)\xi\left(\frac{P}{Q}-\xi\right)
\end{align}
with the shock solution $\xi(-\infty)=0$,  $\xi(\infty)=P/Q$:
\begin{align}
&\xi=A(1+g), \quad A=\frac{P}{2Q}, \quad k=\frac{P}{2S}\end{align}
provided $P/Q>0$, $P/S>0$. 

Ignoring dissipation $S$ one gets
\begin{align}
& R\partial^2_X\xi=-P\xi + Q\xi^2\\
&R(\partial_X\xi)^2+P\xi^2-\frac{2Q\xi^3}{3}=0
\end{align}
with the soliton solution $\xi(-\infty)=\xi(\infty)=0$:
\begin{align}
&\xi=A(1-g^2),\quad 
k=\sqrt{-\frac{P}{R}}, \quad A=-\frac{3P}{2Q}
\end{align}
provided $P/R<0$ and $Q/R<0$. 

In what follows we are particularly interested in the case when $M^2>v_F^2$ so that all parameters $P,Q,R,S$ are positive. Using the scaling transformation 
\begin{align}
&T=\alpha T, \quad X=\beta X, \quad \xi=\gamma\xi\\
&\beta=\sqrt{\frac{R}{P}}=\sqrt{\frac{\cos^2\theta(M^2-v_s^2)}{(M^2-\cos^2\theta)(M^2-v_F^2)(M^2-v_{SL}^2)}} \\
& \alpha=\frac{\beta}{P}=\sqrt{\frac{\cos^2\theta(M^2-v_s^2)^3}{(M^2-\cos^2\theta)(M^2-v_F^2)^3(M^2-v_{SL}^2)^3}}, \\
&\gamma=\frac{P}{Q}=\frac{(M^2-v_F^2)(M^2-v_{SL}^2)}{\sin^2\theta}
\end{align} 
equation \eqref{eq:KdVB1} can be rewritten in the following one-parametric form
\begin{align}
&\partial_T\xi+\partial_X \xi - \partial_X\xi^2+ \partial^3_X\xi - S\partial^2_X\xi=0\label{eq:KdVB2}
\end{align}
where 
\begin{align}
&S=\frac{\mu }{\sqrt{PR}}=\frac{\mu}{\cos\theta}\sqrt{\frac{(M^2-\cos^2\theta)(M^2-v_s^2)}{(M^2-v_F^2)(M^2-v_{SL}^2)}}
\end{align} 
is the  dimensionless parameter which characterizes the dissipation strength. It is significantly Mach number dependent. 
The parameters $\alpha$ and $\beta$ are the typical temporal and spatial scales, while $\gamma$ is the shock amplitude. In the cold case $v_s=0$, $v_{SL}=0$, $v_F=1$, and for weak shocks $M^2=1+\delta$, $\delta \ll 1$, one would have
\begin{align}
&\beta=\frac{\cot\theta}{\delta^{1/2}}, \quad \alpha=\frac{\cot\theta}{\delta^{3/2}}, \quad \gamma=\frac{\delta}{\sin\theta}, \quad S=\frac{\mu\tan\theta}{\delta^{1/2}}
\end{align}
In particular, these expressions show that even for small $\mu$ the effective dissipation parameter $S$ can be large for sufficiently low Mach numbers.

\section{Stationary case}

Stationary solutions satisfy the time-independent equation which, after integration once, gives
\begin{align}
&\xi - \xi^2+ \xi'' - S\xi'=0\label{eq:KdVB2s}
\end{align} 
where we require the $\xi'=0$ and $\xi''=0$ for $\xi=0$. The other point where the derivatives vanish is $\xi=1$.
It was suggested \citep{Jeffrey1991Exact-solutions} that a travelling wave solution may be sought for in the form 
\begin{align}
&\xi=A_1+A_2\tanh [k(X-UT)]+A_3\cosh^{-2}[k(X-UT)]\label{eq:travel}
\end{align}
where the parameters $U$, $A_i$, $i=1,2,3$, should be found by substitution in \eqref{eq:KdVB2s}. Since one can always switch to the frame moving with the travelling wave and re-define/re-normalize all variables, it is equivalent to finding a solution of \eqref{eq:KdVB2s}. 
We, therefore, will try a solution of the form 
\begin{align}
&\xi=\frac{1}{2} (1+g) + A(1-g^2)=(A+\frac{1}{2}) + \frac{g}{2}-Ag^2
\end{align} 
where $g=\tanh kX$.
Upon substitution in \eqref{eq:KdVB2s} one has
\begin{align}
&k=\frac{1}{2\sqrt{6}}, \quad A=-\frac{1}{4}, \quad S=\frac{5\sqrt{6}}{6}
\end{align}
It can be easily seen that $\xi'>0$ for this solution, that is, it describes a monotonic shock transition. 
The  condition on $S$ means that for a given Mach number $M$ the resistivity $\mu$ must have some definite value for this solution to exist.

\section{Shock-like profile}
\begin{figure}
\centering
\includegraphics[width=0.65\textwidth]{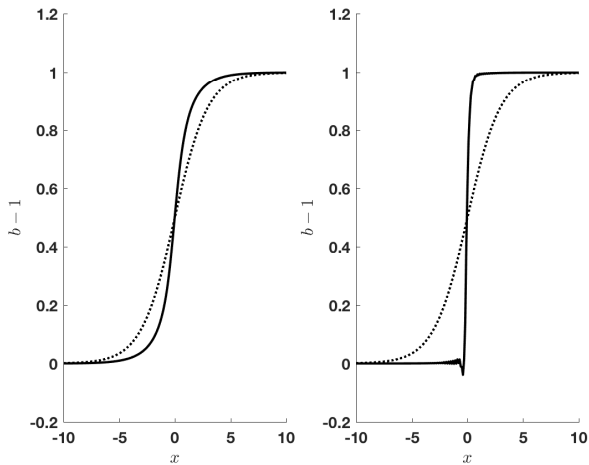}
\caption{Initial (dotted line) and evolved profiles for  $S=1$. }
\label{fig:figs1}
\end{figure}

\begin{figure}
\centering
\includegraphics[width=0.65\textwidth]{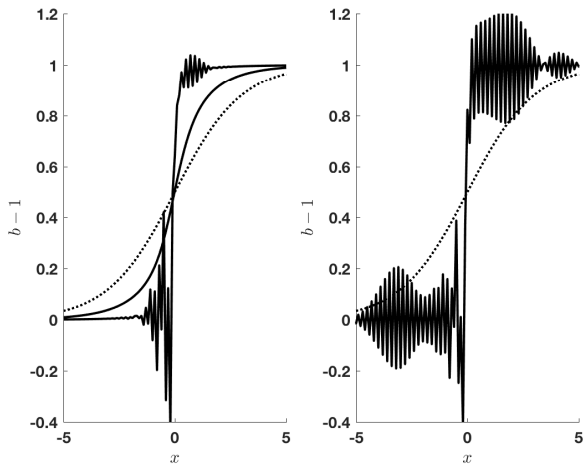}
\caption{Two profiles at two different times for $S=0.25$ and the initial profile (dotted line).}
\label{fig:figs025}
\end{figure}

\begin{figure}
\centering
\includegraphics[width=0.65\textwidth]{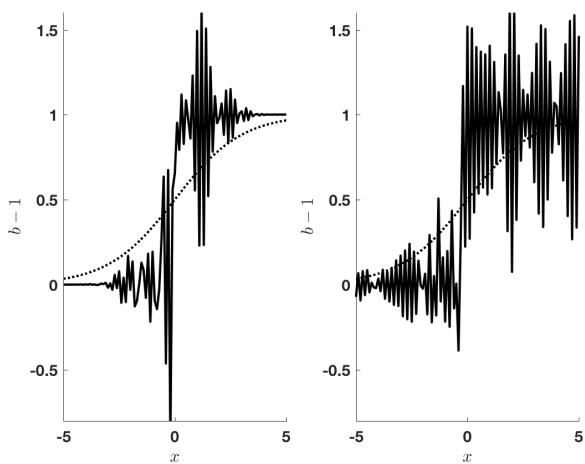}
\caption{Two profiles at two different times for $S=0.005$ and the initial profile (dotted line).}
\label{fig:figs005}
\end{figure}

In the focus of our study is the evolution of an initially smooth profile, of the kind 
\begin{align}
&b=1+\left(\frac{1}{2}\right)\left(\tanh
\left(\frac{x}{D}\right)+1\right)\label{eq:inprofile}
\end{align} 
such that $\xi(-\infty)=b(-\infty)-1=0$, $\xi(\infty)=b(\infty)-1=1$, and $D$ is larger than the typical wavelength. Roughly speaking, this profile describes the transition from one asymptotic value of the magnetic field to another within the distance $D$.  In what follows we present the results of the numerical integration of KdV-Burgers \eqref{eq:KdVB2} for  various values of the dissipation parameter $S$. The integration is done using spatial finite-difference method with the grid spacing $\Delta x=0.1$ and temporal 4th order Runge-Kutta method. Figure~\ref{fig:figs1}
shows two  profiles achieved at two different times as a result of steepening for  $S=1$. The  initial  profile is shown by the dotted  line.  In this case the dissipation is strong enough to ensure quick convergence to a final stationary state, which is not monotonic. This is clearly seen in the movie 1 (online supplementary materials). 
Figure~\ref{fig:figs025}
shows  two profiles for two different times for $S=0.25$. The ramp and the magnetic dip before the ramp are weakly time-dependent. Time-dependent wavetrains are generated both upstream and downstream by the steepened profile.   The variations are illustrated in the movie 2 (online supplementary materials).
Finally, Figure~\ref{fig:figs005}
shows  two profiles achieved at two different times for $S=0.05$. The two late profiles are quite different  which indicates essential non-stationarity of the shock front. The evolution of the profile in this case is shown in the  movie 3 (online supplementary materials). 
It is worth mentioning that although downstream wavetrains are also present, they should not be taken seriously, since ion kinetics is important in this region and the  assumptions of an isotropic polytropic state equation for ions are not valid in this region~\citep{Gedalin2015Collisionless-r}.

\section{Conclusions}
We have shown that weak shocks with weak resistive dissipation can be described by KdV-Burgers equation. This equation is expected to possess stationary solutions. However, it is quite plausible that not any initial smooth profile 
with different asymptotical values of the magnetic field converges to the stationary point or that the convergence rate is extremely slow. In this case the profile will steepen until it begins to generate time-dependent wavetrains which prevent further steepening. Thus, the shock profile is intrinsically time-dependent and stationary only on average. Although the wavetrains are generated in a time-dependent manner, the time-dependence of the ramp (the main magnetic increase) may be weak for a range of dissipation parameter, so that the stationary approximation would be justified for the analysis of the ion motion across the shock front. If the dissipation is strong enough the upstream wavetrain is also nearly stationary or even reduces to a small number of phase standing oscillations adjacent to the ramp. For sufficiently weak dissipation propagating  wavetrains are formed with the changing shape and the shock front experiences reformation.   These findings may explain the observed variety of the  low-Mach number shock profiles~\citep{Farris1993Magnetic-struct,Wilson2007Waves-in-Interp,Russell2009STEREO-observat,Kajdic2012Waves-upstream-,Wilson-III2017Revisiting-the-}. At this stage, there is no good theory providing the dissipation parameters for throughout the shock, including the upstream region. Therefore, no first principle based estimates are available for the control parameter $S$.

This study was supported in part  by the Israel Science Foundation (grant No. 368/14).

%
%

\end{document}